\documentclass[fleqn,10pt]{wlscirep}
\usepackage[utf8]{inputenc}
\usepackage[T1]{fontenc}
\title{The Influence of Ridership Weighting on Targeting and Recovery Strategies for Urban Rail Rapid Transit Systems}

\author[1+]{Aran Chakraborty}
\author[1+]{Yushi Tsukamoto}
\author[2,+]{August Posch}
\author[2,*,+]{Jack R. Watson}
\author[2]{Auroop R. Ganguly}

\affil[1]{Newton South High School, Newton, MA 02459, USA}
\affil[2]{Northeastern University, Boston, MA 02115, USA}
\affil[*]{watson.jac@northeastern.edu}
\affil[+]{these authors contributed equally to this work}

\begin{abstract}
The resilience of urban rapid transit systems to a rapidly evolving threat space is of much concern. Extreme rainfall events are both intensifying and growing more frequent under continuing climate change, exposing transit systems to flooding, while cyber threats and emerging technologies such as unmanned aerial vehicles are exposing such systems to targeted disruptions. An imperative has emerged to model how networked infrastructure systems fail under such scenarios and devise strategies to efficiently recover them from disruptions. Many strategies have been devised, including those based on network centrality measures and passenger flow characteristics. Passenger flow approaches can quantify resilience in more dimensions than network science-based approaches, but the former typically requires large amounts of granular data derived from automatic fare collection and suffer from large algorithmic runtime complexities. Some attempts have been made to include publicly accessible low-resolution ridership data in graph-theoretical frameworks. However, there is yet to be a systematic investigation of the effects of incorporating low-dimensional, coarsely-averaged ridership volumes in strictly topological network science methodologies. We simulate targeted attack and recovery sequences using station-level ridership, centrality measures, and weighted combinations thereof to produce attack and recovery sequences. Resilience is quantified using two topological measures of performance: the node count of a network's giant connected component (GCC), and a new measure we term the "highest ridership connected component." (HRCC). Three transit systems are used as case studies: the subway and light rail systems of the cities of Boston, New York, and Osaka. Simulation results show that centrality-based failure and recovery strategies are most effective when measuring performance via GCC, while centrality-ridership hybrid strategies perform strongest measuring performance by HRCC. The study reveals that the strongest strategies vary based on network characteristics and the kind of functionality most valued by emergency managers and transit operators, highlighting the imperative to plan for the most dangerous attacks and most effective recovery procedures according to each city's unique needs. 

\end{abstract}

\begin{document}

\flushbottom
\maketitle

\thispagestyle{empty}

\section*{Introduction} %
Urban rail transit (URT) networks are vital to human mobility in urban agglomerations, relieving road congestion \cite{anderson2013subways} and providing a comparatively lesser carbon footprint. \cite{fta2010public} URTs provide safe, reliable, high-volume transportation across metropolitan areas and are typically electrified.\cite{uic2015railway} URT resilience has seen heightened interest over the past decade, driven in part by the increasing exposure of URTs to hazards, particularly climate change-exacerbated extreme weather becoming more frequent and intense \cite{Grundemann2022Oct} and cyber-physical attack, which complex infrastructure systems are increasingly exposed to as integrated digital systems become ubiquitous in the operation of infrastructures. \cite{Moller2023}

The resilience of URTs refers to their capacity to withstand and recover from disruptions, failing gracefully and recovering in a timely manner to ensure minimal disruption to service.\cite{Bi2023Jul,Wei2024Mar,Derrible2010Sep} Resilience is a multifaceted paradigm subsuming multiple concepts under its umbrella, typically consisting of robustness, vulnerability, rapidity, and redundancy \cite{Ma2022Dec, Wei2024Mar}. Robustness is the ability of a system to withstand hazards. A robust system is able to fail gracefully, maintaining some level of operational capacity, rather than failing abruptly and totally or in a cascade across components and systems. (@X) Vulnerability, in contrast to robustness, is a system's sensitivity to hazards, or the magnitude of performance disruption due to a given disturbance. Rapidity is the ability of a system to quickly recover from a disruption, returning to normal performance levels. Redundancy is the ability of a system to perform the same function via multiple mechanisms or pathways, such that when one subsystem fails, another redundant subsystem is able to serve the same purpose and "pick up the slack". It is worth noting that resilience and its constituent aspects lack meaning without a prepositional modifier: a subway system is resilient \textit{to} coastal flooding; a fuel transportation network is vulnerable \textit{to} cyberattacks. The exception is redundancy, which is an inherent property of a system. Regardless of any particular hazard, a system may or may not possess redundancies in its functional pathways. However, this is not to say that redundancies are not designed (or evolved) in anticipation or response to certain kinds of hazards.

The failure dynamics of URTs subject to disruption is composed of complex processes, involving train schedules and operational changes, human mobility patterns, and transit network structure, along with interdependencies between power grids, telecommunications, and other modes of transportation. \cite{Lu2019Apr, Ilalokhoin2023Jan, johansson2011vulnerability, Adjetey-Bahun2016Sep} Consequently, the modeling and simulation of failure is a complex undertaking, with myriad system representations each with their own simplifying assumptions. Approach is often dictated by data availability, with granular models of physical infrastructure and network flows being in many cases impractical or unfeasible. Therefore simplified network representations are attractive due to more lenient data requirements and computationally efficient approximations of network optimization algorithms. A major trade-off between demanding, intensive microscopic models and lower-dimensional network science approaches is that the robustness of URTs to perturbations can be highly sensitive to the state of traffic within the system. \cite{Tang2021Oct} The measure of performance chosen is largely a function of objective: when physical infrastructure integrity is of interest, the size of the network's giant connected component (GCC) is often used as a proxy for system functionality. \cite{Bhatia2015Nov, Watson2022Nov, Yadav2020Jun} When considering traffic flow, total trip delay compared to a normative baseline commute pattern, or demand impedance, is used to quantify performance. \cite{Chen2022Jan, Tang2021Oct, Sun2018Feb, Fan2020Mar}

When passenger flow data is available, simulations typically takes the form of dynamic traffic network simulations using historical passenger flow data inferred via automatic fare collection (AFC) \cite{Wang2013Nov, Sun2015May, Sun2018Feb, Lu2019Apr, De-Los-Santos2012Feb, Xu2022Jul}. A limitation to this approach is the often proprietary status of such datasets, access to which may be controlled via institutional partnerships between research institutions and transit agencies. In many cases, publicly available data on URT traffic flow is limited to average ridership on a station (or component, more generally) level, leading modelers to either generate synthetic traffic flow inferred from  other data sources or limit measures of resilience to topological proxies like GCC size. The former includes inferences using cellular GPS human mobility data \cite{Wang2013Nov} and face-to-face interviews \cite{Deng2015Dec}. Several optimization approaches have also been implemented: Tang et al. \cite{Tang2021Oct} developed a linear programming optimization model, with disruptions causing rerouting of passenger flow; Fan et al. \cite{Fan2020Mar} developed a \textit{temporal subway network} model in which edge weights, representing passenger flows, change over time on a fixed network topology; and Chen et al. \cite{Chen2022Jan} proposes a bi-level optimization model to minimize passenger accessibility and maximize network efficiency. In the absence of ridership data, topological proxies are often employed.\cite{Yadav2020Jun, Watson2022Nov, Bhatia2015Nov} 

A compromise between the two is the use of average ridership to weight failure and recovery sequence strategies (@X). This is a convenient modeling assumption because it allows for the simplicity of a strictly topological simulation approach while factoring in ridership without constructing and solving a dynamical systems and/or network flow model (@X). However, it is unclear how suitable this approach is, and to what degree, if any, added ridership information improves the performance of failure and recovery strategies. To our knowledge, this study constitutes the first systematic investigation of the effects of weighting topology-based failure and recovery strategies with average station-level ridership volume. 

This investigation is concerned with the ability of URTs to (1) absorb disruptions, failing gracefully rather than collapsing suddenly, and (2) recover quickly and efficiently, bringing system functionality back to normal levels in a timely manner. We ask the following questions: do targeted attack strategies become more effective-- that is, increase the vulnerability of the targeted system-- by accounting for average ridership when determining station deactivation sequences? Or, framed another way, does accounting for ridership during targeting decrease the robustness of the network? Second, does accounting for average ridership increase the rapidity of recovery, leading to component recovery sequences that more efficiently return the system to normal functionality? If so, why? What combination of topological network centrality-based strategies and ridership-based strategies leads to the highest resilience? Does changing the topological measure of system functionality change the optimal failure and recovery strategies? What do different functionality measures tell us about the state of the URT systems of interest? What can be generalized to other URTs, and other kinds of networks? 

\begin{figure*}
  \includegraphics[width=\textwidth]{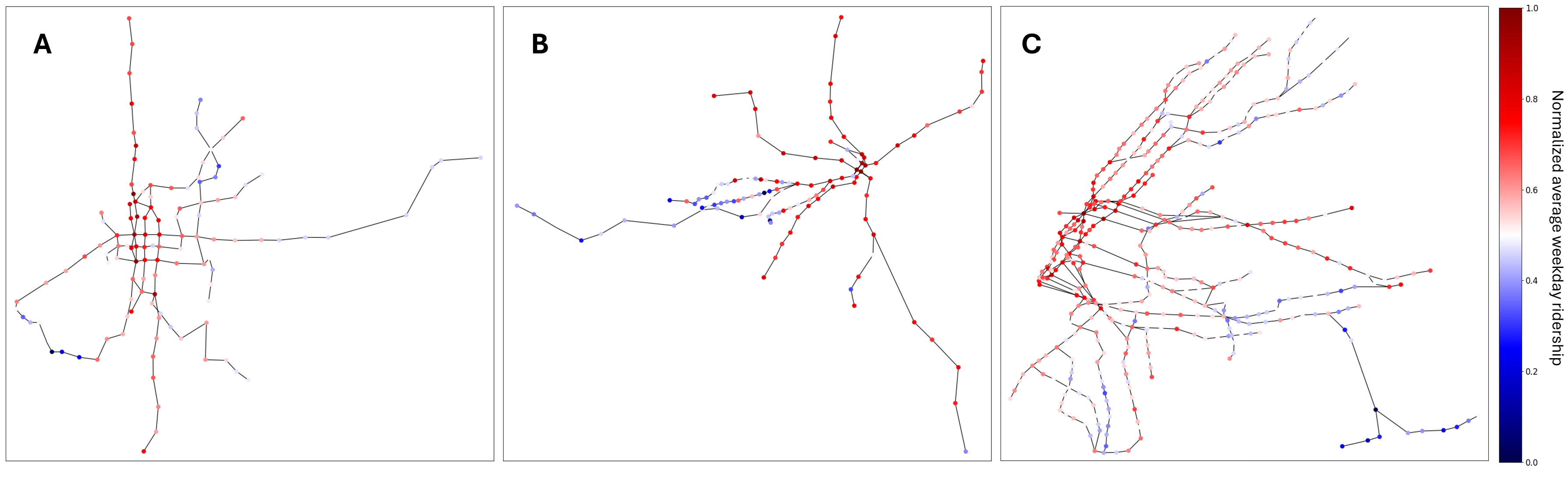}
  \caption{Transit system network topologies constructed in L-space. Nodes represent stations, with node positions corresponding to the latitude and longitude of stations. Node color corresponds to the average weekday ridership per station, rescaled to range [0,1]. Edges represent the structural adjacency between stations and are undirected (or bidirectional). A: Osaka Metro. B: Boston MBTA Subway ("The T"). C: New York City MTA Subway.}
  \label{transitnetworks}
\end{figure*}

\section*{Results} %

\begin{figure*}
  \includegraphics[width=\textwidth]{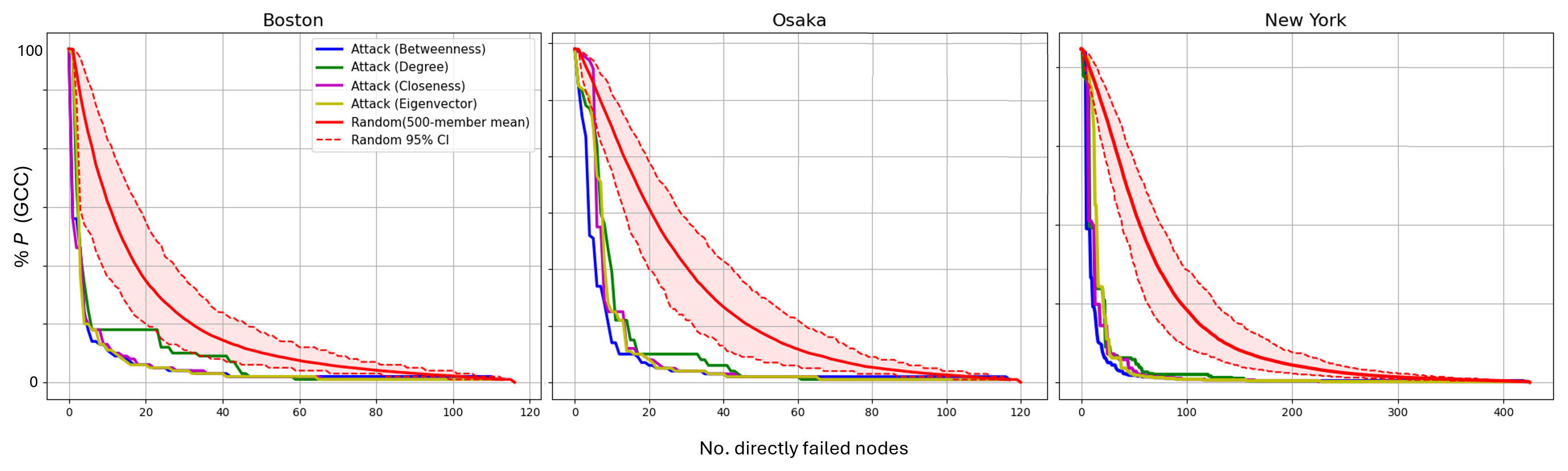}
  \caption{Resilience curves measured by GCC (highest nodes in connected body) under attack by four centrality-based targeting strategies (Betweenness, Degree, Closeness, Eigenvector) and the mean of 500 random simulation curves with a 95\% confidence interval. From left to right: Boston T, Osaka Metro, New York City Subway.}
  \label{curvesattackcentralitygcc}
\end{figure*}

\begin{figure*}
  \includegraphics[width=\textwidth]{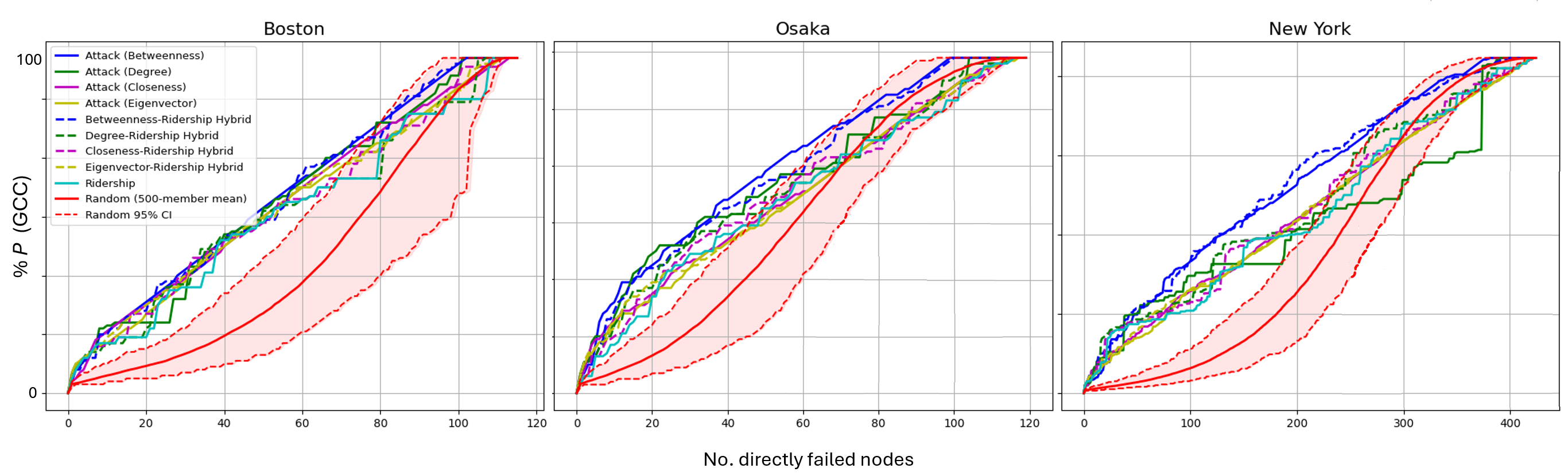}
  \caption{Resilience curves measured by GCC (highest nodes in connected body) under recovery by four centrality-based recovery strategies (Betweenness, Degree, Closeness, Eigenvector), ridership-based recovery, four hybrid strategies weighting ridership with each aforementioned centrality (Betweenness-Hybrid, Degree-Hybrid, Closeness-Hybrid, Eigenvector-Hybrid), and the mean of 500 random simulation curves with a 95\% confidence interval. From left to right: Boston T, Osaka Metro, New York City Subway.}
  \label{curvesrecoverygcc}
\end{figure*}

\begin{figure*}
  \includegraphics[width=\textwidth]{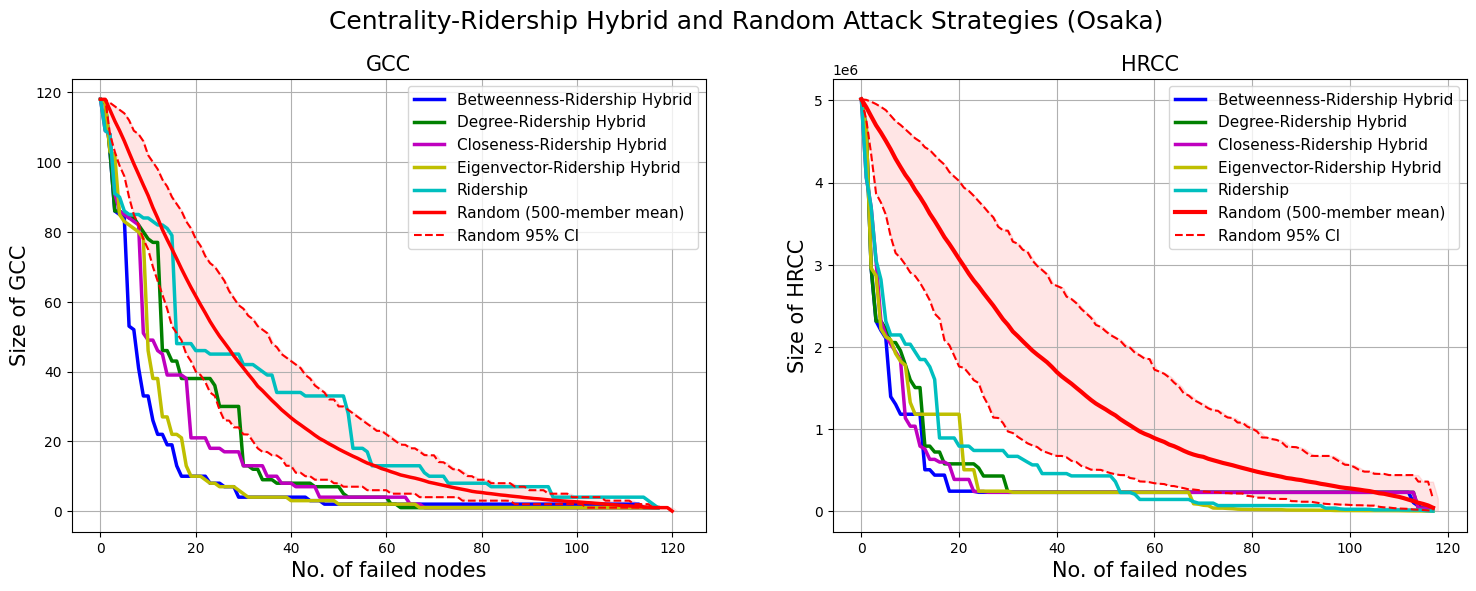}
  \caption{Resilience curves for Osaka Metro network when under attack by ridership-based targeting, four hybrid strategies weighting ridership with each centrality (Betweenness-Hybrid, Degree-Hybrid, Closeness-Hybrid, Eigenvector-Hybrid), and the mean of 500 random simulation curves with a 95\% confidence interval. Left measures GCC (highest nodes in connected body), right measures HRCC (highest ridership in connected body).}
  \label{curvesattackhybridosaka}
\end{figure*}

\begin{figure*}
  \includegraphics[width=\textwidth]{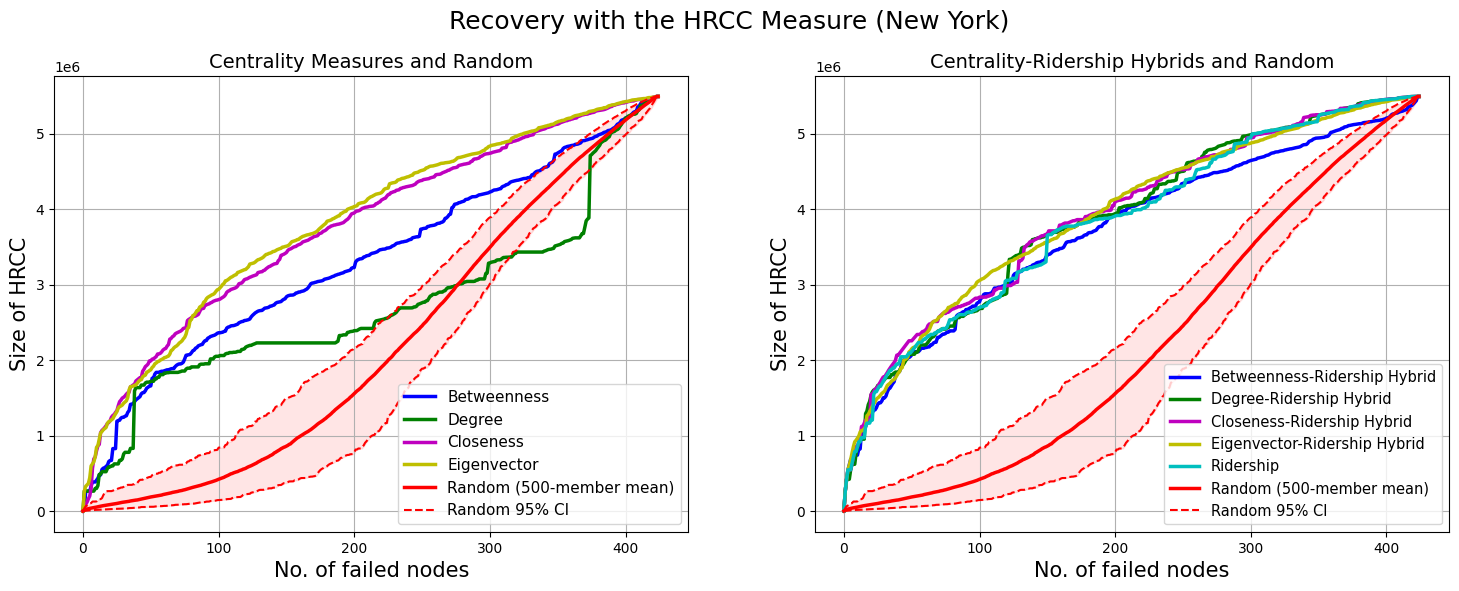}
  \caption{Resilience curves for New York City Subway network by HRCC (highest ridership in connected body) during recovery phase. Left graph shows curves for four centrality-based attacking strategies (Betweenness, Degree, Closeness, Eigenvector), and right graph shows curves for ridership-based targeting, as well as four hybrid strategies weighting ridership with each centrality (Betweenness-Hybrid, Degree-Hybrid, Closeness-Hybrid, Eigenvector-Hybrid). Both graphs show the mean of 500 random simulated curves with a 95\% confidence interval.}
  \label{curvesrecoveryhrccnewyork}
\end{figure*}

\begin{figure*}
  \includegraphics[width=\textwidth]{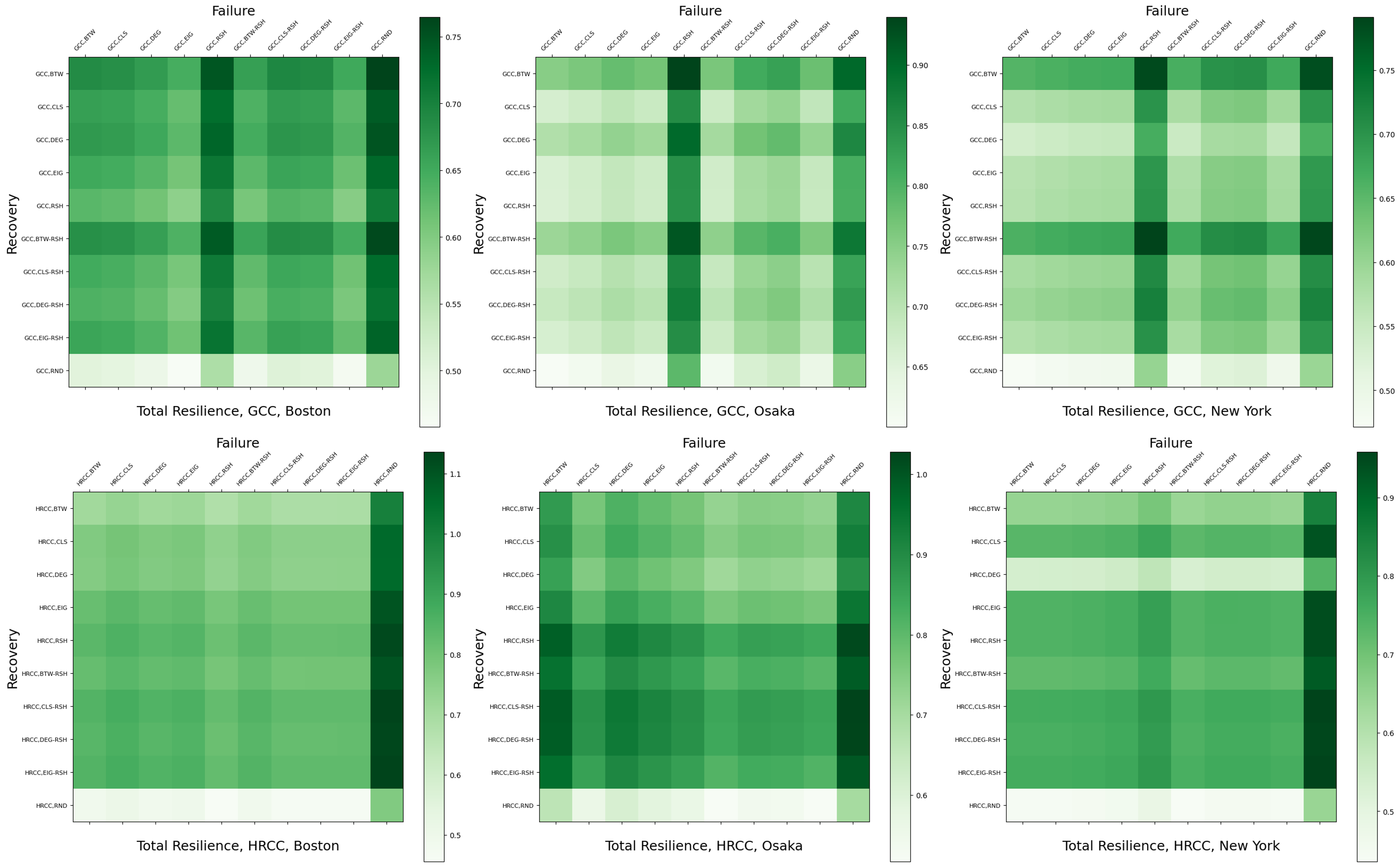}
  \caption{Heatmaps comparing the overall measured resilience of every situation per city, where in each situation, one of the attack strategies is met with one of the recovery strategies. The strategies ordered from top to bottom for recovery and left to right for attack are as follows: Betweenness Centrality, Degree Centrality, Closeness Centrality, Eigenvector Centrality, Ridership, Betweenness-Ridership Hybrid, Degree-Ridership Hybrid, Closeness-Ridership Hybrid, Eigenvector-Ridership Hybrid, Random Simulation Mean Curve. The top row of heatmaps measure by GCC (highest nodes in connected body); the bottom row of heatmaps measures by HRCC (highest ridership in connected body). Each column corresponds to a city from left to right: Boston T, Osaka Metro, New York City Subway.}
  \label{heatmaps}
\end{figure*}

\begin{figure*}
  \includegraphics[width=\textwidth]{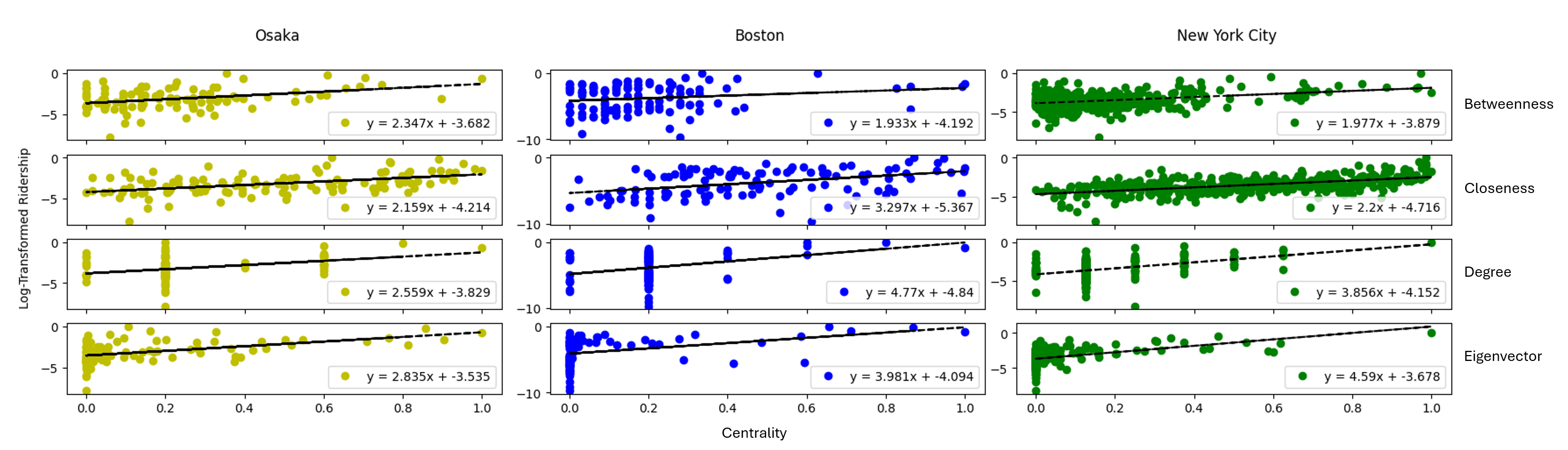}
  \caption{Linear correlation between centrality scores and log-transformed average weekday ridership for all stations within each transit system. From top to bottom rows: betweenness, closeness, degree, and eigenvector centrality. From left to right: Osaka Metro (yellow), Boston T (blue), and New York Subway (green).}
  \label{ridershipcentralitycorr}
\end{figure*}

To investigate the strength of topological and ridership-based strategies, we performed attack and recovery simulations on the urban rail transit networks of Boston, Osaka, and New York, and evaluated the strategies with a topology-based measure of functionality as well as functionality measure including ridership. The topological strategies included the purely centrality based betweenness, degree, closeness, and closeness, and eigenvector centrality strategies. A ridership strategy ordered the stations based on their typical ridership. Lastly, ridership-centrality hybrid strategies created an ordering of stations based on both ridership and their centrality. The strategies were compared against a 500-member random ensemble of simulations in which the attack or recovery targeted stations in a random order. Finally, resilience scores were given to each strategy based on the network functionality throughout the simulation. One measure of functionality was the size of the giant connected component (GCC), which counts the number of contiguously connected stations at each step of the simulation. The other measure of functionality was the ridership sum in the highest ridership connected component (HRCC), which identifies the highest-ridership contiguously-connected set of stations at each step of the simulation. Once the Boston, Osaka, and New York transit systems were represented as graphical networks, the simulations of all strategies were performed and strongest strategies were identified.

\subsection*{Targeted attack strategies}
In two out of three cities, the strongest attack strategy depended on the choice of functionality measure, GCC or HRCC. New York City was the exception, where betweenness centrality was the strongest attack strategy independent of how functionality was measured. However, for Boston and Osaka the strongest attack strategy depended on how functionality was measured. 

When using GCC as a functionality measure, the strongest attack strategy in Boston was eigenvector centrality, closely followed by the eigenvector-ridership hybrid strategy. For Osaka and New York the strongest attack strategy was betweenness centrality. The centrality-based strategies are featured in Figure \ref{curvesattackcentralitygcc} shows the strongest strategies as well as how similar the resilience curves are shaped for all three networks.

When using HRCC as a functionality measure, the strongest attack strategy in Boston was the ridership strategy. For Osaka, the most effective strategy was the betweenness-ridership hybrid strategy, followed closely by the eigenvector-ridership hybrid strategy. For New York City, the strongest attack strategy was betweenness centrality, and in general, purely centrality-based strategies outperformed ridership-based strategies for New York's network.

\subsection*{Recovery strategies}
In all three cities, the strongest recovery strategy depended on the choice of functionality measure. 

When measuring functionality as the number of nodes contained within the giant connected component (GCC), betweenness centrality was found to be the optimal recovery strategy across all three transit systems. This is in agreement with previous studies (@X cite). 

When measuring functionality by HRCC, the optimal recovery strategy for both Osaka and Boston was closeness-ridership hybrid strategy. For New York City, the optimal recovery strategy was closeness centrality. Strategies resulting in higher resilience scores for New York City, as measured by HRCC, were consistently divergent from those observed for Osaka and Boston. Purely centrality-based strategies outperformed strategies that included ridership. 

\subsection*{Summary of results}
Certain kinds of strategies had their best use in certain scenarios. Centrality-ridership hybrid strategies were generally strongest when HRCC was the functionality measure, while purely centrality-based strategies were generally strongest when GCC was the functionality measure. This is visible in the heat maps of resilience by strategy combination in Figure \ref{heatmaps}. Across cities and attack/recovery and measures of functionality, choosing the strongest strategy always outperformed the random mean, although not always outperformed the 97.5 percentile random simulation; however, choosing a strategy that was in the middle or worse end of the spectrum for a given scenario was often worse than the random ensemble mean. 

\begin{figure*}
  \includegraphics[width=\textwidth]{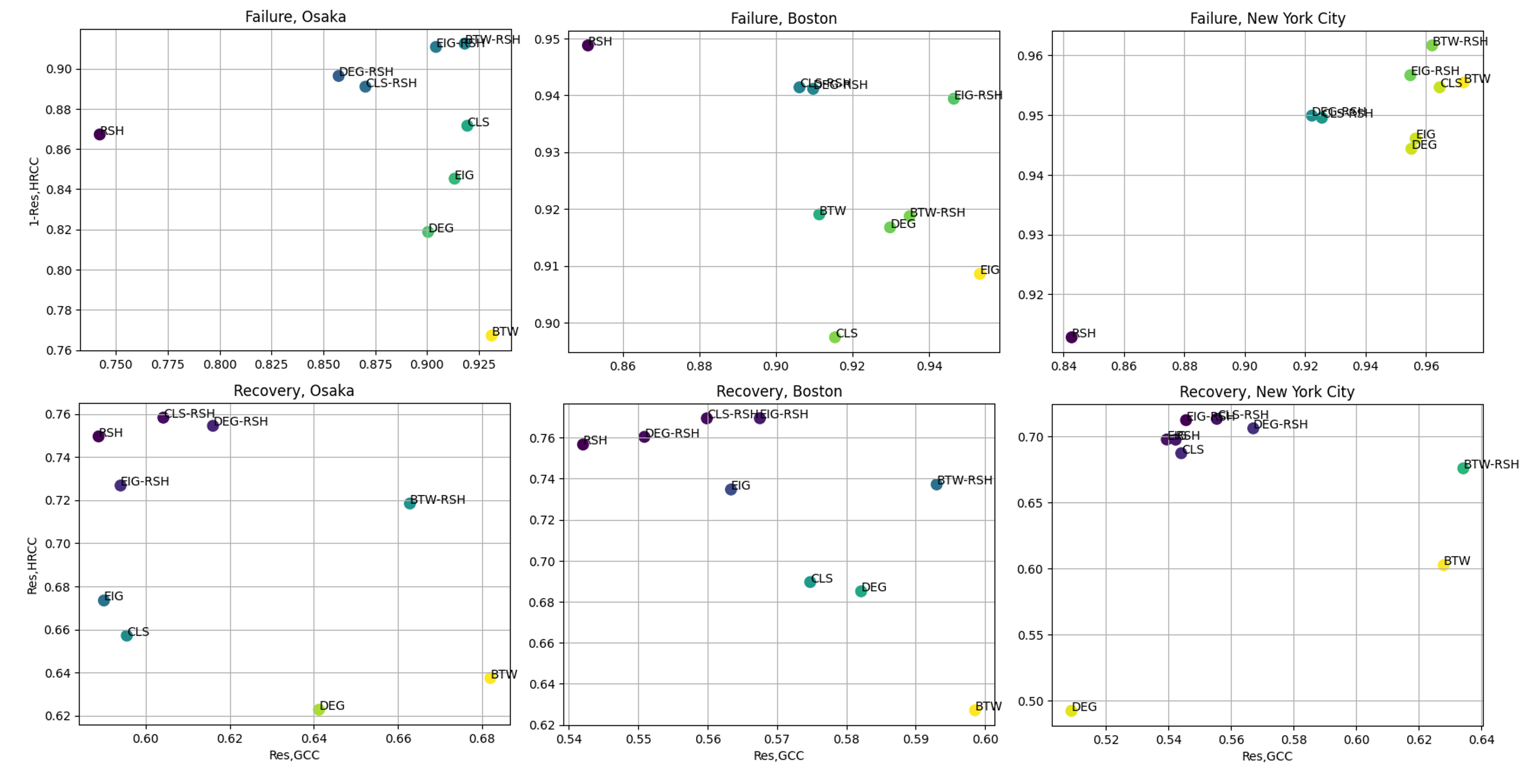}
  \caption{Total resilience scores for failure (robustness) and recovery simulations. The resilience of each failure and recovery strategy is plotted using both HRCC and GCC as a performance metric. Failure is plotted as $1 - R_{tot}$ to preserve the visual intuition that larger scores correspond to better performance as holds for recovery. This is because the more effective a targeted attack strategy is at degrading performance, the lower the robustness of the network.}
  \label{scatterplots}
\end{figure*}

For all three cities in attack and for Boston and Osaka in recovery, the strategy that was strongest as measured by one functionality metric was middling or weak as measured by the other functionality metric. For example, in Boston attack, ridership is the strongest strategy by HRCC but weakest by GCC, and eigenvector is the strongest strategy by GCC but second-weakest by HRCC. This is visible in Figure \ref{scatterplots}. Similar trade-offs appear in most other scenarios, although less extreme. Across attack/recovery and functionality measures, the choice of strategy mattered more in New York and less in Boston. This was measured by calculating the variance of resilience scores across all of a city's scenarios.

Different cities and functionality measures also featured different shapes of their best strategy's resilience curve. An example of this is New York City recovery, as its strongest strategy by GCC has a mostly linear shape (Figure \ref{curvesrecoverygcc}), while its strongest strategy by HRCC has a roughly logarithmic shape (Figure \ref{curvesrecoveryhrccnewyork}). The same difference in curve shape held true for Boston, to an even greater extent (Figure \ref{curvesrecoverygcc} and Supplementary Figure 1). In addition, there could be great variance in curve shapes even among the same types of strategies, shown in the centrality-ridership hybrid strategies for attack in Osaka (Figure \ref{curvesattackhybridosaka}), which are disparate whether measuring by GCC or HRCC.

\section*{Discussion} %
\ \

In general, the strongest strategy depends on the choice of functionality measure, and this is true across cities and across attack and recovery scenarios; the notable exception is New York City attack, whose strongest strategy was betweenness regardless of choice of functionality measure. Because different functionality measures result in different optimal strategies, emergency managers must carefully consider what kind of functionality they care about. (Note, the stakes are even higher in Boston and Osaka because the strategy that was strongest as measured by one functionality measure was near-weakest as measured by the other functionality measure.) If emergency managers care about keeping the highest-traffic areas functional in the event of a hazard - or, more specifically, keeping some contiguous set of stations with highest traffic functional - then HRCC is the appropriate way to evaluate the most dangerous attacks and plan in advance for the most efficient recovery strategies. If, on the other hand, emergency managers care about keeping the greatest number of stations contiguously connected, then GCC is the appropriate way to plan for attacks and develop recovery strategies. Emergency managers need to make this decision based what is most relevant to recovery in their network, which may also depend on how lifeline services such as hospitals and airports, economic drivers such as employment centers, or socially vulnerable populations such as those who do not own a car or live below the poverty line, are distributed throughout their urban rail network. It would also be possible to develop, beyond GCC and HRCC, a new measure of system functionality based on the unique considerations of a city and its desired emergency response. However, already we can see that it is possible to evaluate strategies based on a blend of GCC and HRCC. Looking at the scatter plots (Figure \ref{scatterplots}), the strongest strategies by GCC are the the rightmost points and the strongest strategies by HRCC are the uppermost; emergency managers could select the strategies that are furthest to the upper-right if a blend of GCC and HRCC is desired. For example, for Boston attack the eigenvector-ridership hybrid strategy is strongest if an equal blend of GCC and HRCC is what practitioners care about. The goal of this study is not to tell emergency managers which measure to care about, but rather to reveal that the optimal strategy depends on how functionality is measured.

Aside from discovering the strongest strategy in each scenario, it is also insightful to look at the problem in reverse and identify which scenarios are the best fit for each strategy. Indeed, this study found certain sets of strategies are most useful in certain contexts. For example, this study is among the first to thoroughly evaluate centrality-ridership hybrid strategies. In most scenarios measured by HRCC, including all three cities for recovery and Osaka and New York for attack, the strongest strategy was one of the centrality-ridership hybrids. In the only other scenario, Boston attack, pure ridership was the strongest, revealing that in all scenarios tested, some ridership-influence was required to be the strongest strategy by HRCC. But in five of six scenarios, the pure ridership strategy performed worse than the centrality-ridership hybrid strategies when measuring functionality by HRCC - a measure that explicitly takes into account ridership. This interesting dynamic may be a feature of how HRCC is defined - it is the highest-ridership \textit{connected} component. So recovering or attacking high-ridership stations in disparate parts of the network may be an ineffective strategy for affecting that key connected component, while the centrality part of the centrality-ridership hybrid may tend to select station nodes that are connected to the important component. This could be investigated further in future research.

Another example of certain strategies best in certain contexts is the result that purely centrality-based strategies tended to be strongest when functionality is measured by GCC. Just as it is unsurprising (@X change wording) that strategies involving ridership information perform well by HRCC, it is also unsurprising that centrality-dominated strategies perform well by GCC. GCC is the giant connected component, which counts the greatest number of connected stations without regard for ridership. Intuitively, topology-based centrality measures (betweenness, eigenvector, closeness, degree), which do not account for ridership, may make for strong strategies when measuring functionality this way. The experimental results of this study confirm those intuitions when applied to Boston, Osaka and New York urban rail networks.

The fact that betweenness centrality was an optimal recovery strategy for all three systems when using GCC as a functionality measure is interesting. This may be the case because the high-betweenness nodes in the network tend to be located at @X

The choice of strategy matters most in NYC and matters least in Boston, as measured by the range of resilience from the strongest to weakest nonrandom strategy under a given functionality metric. New York City's range was around 0.20 for all its scenarios, whereas Boston's range was around 0.10 for most of its scenarios. The variation among strategies is visible in Figure \ref{curvesrecoverygcc}, where by GCC the recovery curves for all of Boston's strategies are extremely similar, while New York City has a lot of variance with different shapes as well. This dynamic may be related to the different network characteristics of Boston and New York City, where in Boston all the lines radiate from one central hub, while in New York there is a vast number of connections.

The shape of the resilience curve differs across cities and functionality measures. For example, the strongest strategy for New York recovery by GCC has a mostly linear shape, while the strongest strategy for New York recovery by HRCC has a roughly logarithmic shape. This may be due to many possible reasons. A nonlinear shape in attack (steep drop off early) means that an attacker can hamper most of the system functionality in just a few moves, while a nonlinear shape in recovery (steep rise early) means that emergency managers can recover most of the network functionality by rehabilitating just the first few stations.

New York was different than the other cities in notable ways: its strongest strategy for recovery did not depend on the choice of functionality measure (betweenness was the strongest recovery either way), and switching strategies had the greatest effect on resilience, as measured by the variance in resilience across strategies. The shape of the curves by HRCC (Figure \ref{curvesrecoveryhrccnewyork}) reveals part of the explanation, as the degree and betweenness strategies were clearly weaker than other strategies during the middle to late stages of recovery; those two strategies were low-end outliers, such that switching strategies provides practitioners much more resilience. These strategies being so much weaker may also be due to unique aspects of New York's network, such as the high number of stations and the distribution of ridership around those stations, e.g., high ridership existing in a part of the network that degree centrality struggles to find.

\section*{Methods} %

\subsection*{Data}

\subsubsection*{Transit network topology}
Transit network topological structure was represented as an undirected, unweighted adjacency matrix. Adjacency matrices were constructed manually using a combination of publicly available station location data and official transit system maps. Transit systems were represented in L-Space.

\textit{Boston}: Station locations for the MBTA's rapid transit system were obtained from the Commonwealth of Massachusetts MassGIS database. \cite{massgis_mbta_rapid_transit} The adjacency of the network was constructed by hand using a system map of the T. The Mattapan Trolley (or Mattapan Line), a streetcar service extending south from the Ashmont station terminus of the red line and consisting of seven stations (Cedar Grove, Butler, Milton, Central Avenue, Valley Road, Capen Street, and Mattapan) was excluded from this analysis due to the unavailability of average weekday ridership data. 

\textit{Osaka}: Station locations for the Osaka Metro were sourced from Rosenzu.net, a privately-maintained, publicly-available database of Japanese rail system information.\cite{rosenzu_kansai} Adjacency was manually constructed using the official system map of the Osaka Metro. \cite{osaka_metro_route_map}

\textit{New York City}: Station locations for the New York City MTA Subway system were obtained from the US Data.gov database. \cite{mta_subway_stations} The adjacency matrix defining the network topology of the subway was constructed manually using the station data and the official MTA subway map as a reference. Unique among the three subway systems of study, New York has several station complexes. Station complexes are concourses connecting officially discrete stations via tunnels and walkways, with fare gates granting access to all stations within a complex. Because station complexes collect fare at a single point corresponding to multiple stations, ridership data is aggregated at the station complex level. Accordingly, we represent each station complex as a single node with a single ridership value representing all stations within the complex. The official station count for the New York subway is 472 stations; however, when station complexes are aggregated, the network is composed of 423 nodes representing stations and station complexes. 

\subsubsection*{Average weekday ridership}
Station-level ridership data was procured for our three networks for Boston, Osaka, and New York respectively. Preprocessing routines for each ridership dataset are described below.

\textit{Boston}: For Boston's T rapid transit system, station-level average weekday ridership was only available as an average over the fall of 2019. To correct for potential seasonal differences in ridership, a second dataset was utilized, presenting monthly average ridership for all months aggregated by transit line (red, blue, orange, and green lines). First, we compute seasonal averages for line-level ridership. If we let $w_{i}$ be the average weekday ridership for station $i$ across all seasons, then

\begin{align}
w_{i} = r^{i}_{S,F} \frac{r^{I}_{L,F} + r^{I}_{L,Sp} + r^{I}_{L,Su} + r^{I}_{L,W}}{4r^{I}_{L,F}}
\end{align}

where $r^{i}_{S,F}$ is average weekday ridership for station $i$ for the fall season, $r^{I}_{L,F}$ is average weekday ridership for transit line $I$ during the fall (where subscripts Sp, Su, and W represent Spring, Summer, and Winter, respectively), and $i \in I$, i.e., station $i$ is a member of line $I$. This provides a correction factor adjusting for differences in fall ridership from ridership across all seasons. All data was collected from the MBTA Blue Book Open Data Portal.

\textit{Osaka}: The ridership data was obtained from “Osaka Metro Co.”, the official Osaka Metro website. These figures are from a November 12th, 2019, a Tuesday. The Osaka Metro only measures ridership on one day each year, so 2019 was chosen to avoid pandemic effects and the November 12th values are assumed to be typical of average weekday ridership on Osaka's urban rail system. This is a large assumption but a necessary one due to data availability, and the single-day data are still extremely useful for this study's analysis, as these numbers are meant to give a rough estimate of how much people/passengers are relying on this infrastructure and the population being impacted when an attack or hazard occurs. Some stations are part of complexes, where ridership is counted for the entire complex; in these cases, the ridership of the complex was divided equally among all stations of the complex in order to have a single ridership number for each station.

\textit{New York City}: The ridership data for the New York City Subway was acquired from the official Metropolitan Transportation Authority (MTA) website. The dataset provides average annual weekday ridership every year from 2017 to 2022 for 423 stops, including several station complexes.

\subsection*{Strategies for attack and recovery}
We perform attack and recovery simulations on the Boston, Osaka, and New York City subway systems. For both attack and recovery, we use a variety of strategies, including centrality-based strategies (betweenness, eigenvector, closeness, degree), a ridership-based strategy, and ridership-centrality hybrid strategies that blend the importance of centrality and ridership.

Each strategy is defined by an ordering of stations: for example, using the ridership strategy for attack means the highest-ridership station is disabled first, the second-highest-ridership station is disabled second, and so on until all stations have been disabled. Likewise, using the ridership strategy for recovery means the highest-ridership station is recovered first, the second-highest-ridership station is recovered second, and so on until all stations have been recovered. Each strategy is described in detail below.

\subsubsection*{Ridership-based strategy}
As described above, the ridership strategy defines the ordering of stations based purely on the ridership values of the stations.

\subsubsection*{Centrality-based strategies}
Centrality measures were used for simulating attack and recovery. A centrality measure in a network are quantitative metrics used to identify the most important or influential nodes within a network. There are numerous centrality measures, but Degree, Betweenness, Closeness, and Eigenvector are crucial for transportation networks as they effectively identify key hubs, control points, accessibility, and influential nodes in the network. In an attack simulation, the strategy determines how to disconnect nodes in an order from larger to least centrality levels. If nodes with high centrality are attacked, not only are the targeted nodes destroyed but also network lines are isolated and prevented from operating effectively. However, if low centrality nodes were attacked first, only the targeted nodes would be destroyed; the entire network wouldn’t be affected. Similarly for recovery, if the nodes with higher centrality are recovered first, the functionality of the entire network increases.

Betweenness centrality: a measure of centrality determined based on the number of shortest paths that go through the node of interest. \cite{Freeman1977}
\begin{align}
C_{B}(v) = \sum_{s \neq v \neq t \in V} \frac{\sigma_{s,t}(v)}{\sigma_{s,t}}
\end{align}

Degree centrality: a measure of centrality identified by the number of edges. (neighboring nodes) \cite{Hakimi1962}
\begin{align}
C_{D}(v) = \text{deg}(v)
\end{align}

Closeness centrality: a measure of centrality indicating how close nodes were from the entire network.\cite{Bavelas1950}
\begin{align}
C_{C}(v) = \frac{V-1}{\sum_{u}d(u,v)}
\end{align}

Eigenvector centrality: a centrality measured by its influence levels, nodes neighbored with important ones tended to have higher eigenvector measure. \cite{Bonacich2007}
\begin{align}
C_{E}(v) = \frac{1}{\lambda} \sum_{u \in M(v)}{C_{E}(u)} = \frac{1}{\lambda} \sum_{u \in V} a_{v,u} C_{E}(u)
\end{align}

\begin{figure*}
  \includegraphics[width=\textwidth]{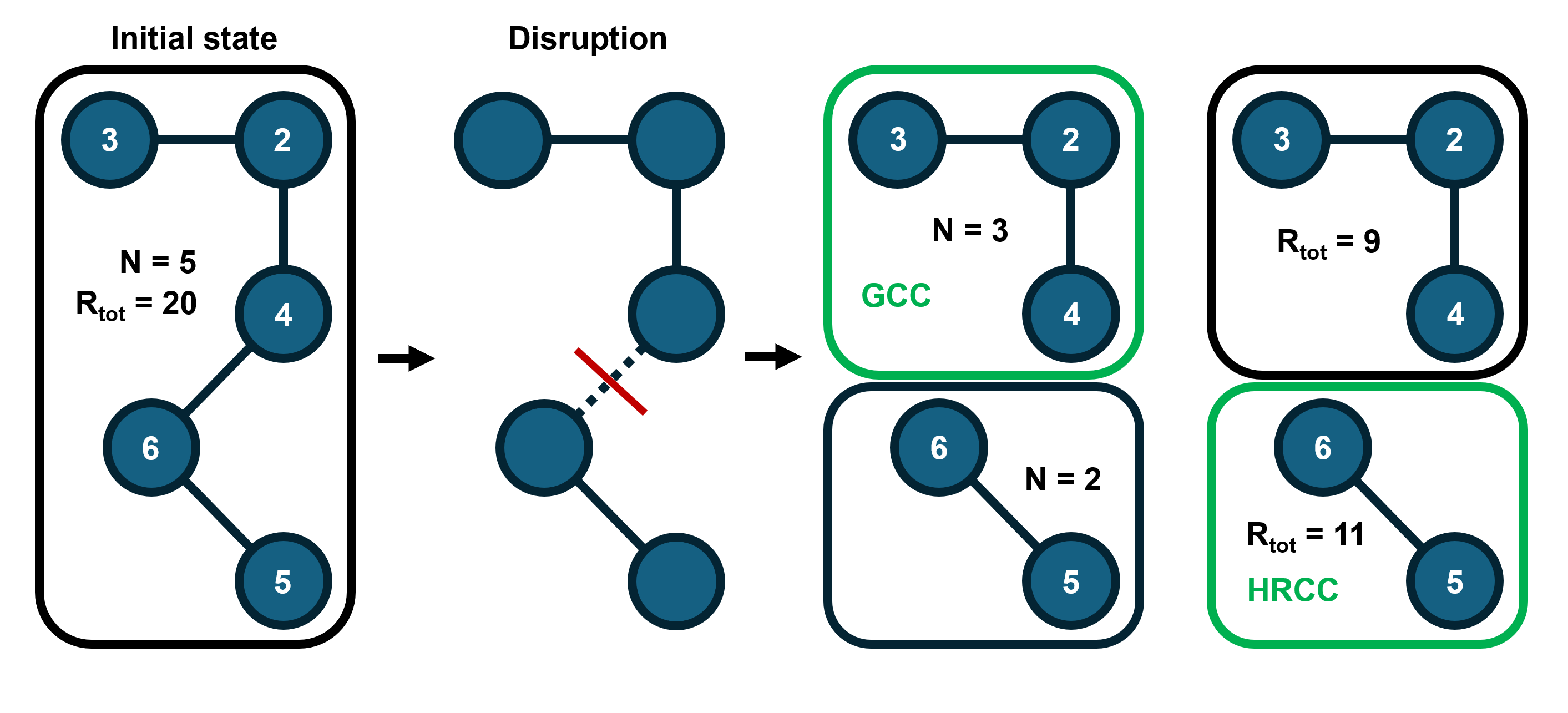}
  \caption{Demonstration of the two performance measures used: the size of the giant connected component (GCC) and the highest ridership connected component (HRCC). The GCC is the connected, or mutually reachable, component in a network with the highest node count. HRCC is the connected component in the network with the highest total ridership when the ridership of each station (node) in the component is summed.}
\end{figure*}

\subsubsection*{Hybrid Strategies}
So far we've described two classes of failure and recovery strategies: centrality-based and ridership-based. We also explored a third strategy combining the two; as centrality and ridership weights are normalized between 0 and 1, the two were multiplied together to weight equally based on centrality and ridership. It was then determined if weighting by combined centrality-ridership resulted in changes in the optimal centrality strategy for failure and recovery. 
\begin{align}
C_{H}(v) = \frac{C_{k} + C_{R}}{2}
\end{align}

\subsubsection*{Random Strategies}
Other than four different centrality-based strategies, simluations were also performed where nodes were attacked/recovered in a random order. To find the 500 member random mean, 500 random simulations were performed and the average was taken to find out the random mean resilience value. In addition to the random mean, a 95\% Confidence Interval was also defined. The 95\% confidence interval was the range within the middle 95\% of multiple random simulations. The upper bound of the 95\% confidence interval is the 97.5 percentile because it was better than 97.5\% of all the random simulations. In other words, only a small percentage had better simulations and the majority of them had worse simulations. Contrarily, the lower bound of the 95\% confidence interval is the 2.5 percentile because it was better than only 2.5\% of all the random simulations, meaning most of the simulations were better than in this bound and very few simulations were worse than this bound.

\subsection*{Measuring system functionality}
Two ways to measure overall system functionality were selected: (1) the size of the giant connected component (GCC) and (2) the sum of station-level average ridership within the giant connected component, labeled here as Highest Ridership Connected Component (HRCC).

Following \cite{Watson2022Nov, Wei2024Mar, Yadav2020Jun}, resilience is calculated as the time-integrated total system functionality over the period of disruption, normalized by the integral of normal functionality over the same duration:

\begin{align}
R = \frac{\int_{t_{0}}^{t_{f}} P(t)dt}{P_{0}(t_{f} - t_{0})}
\end{align}

\subsection*{Experiments performed}
Simulations were performed for every combination of attack/recovery, strategy, and rail network; for each simulation the functionality of the system was measured both by GCC and by HRCC. Resulting from all these simulations we have @X resilience curves.

Resilience curves and resilience curves were evaluated to identify the strongest attack and recovery strategies and understand differences across networks, strategies, and measures of functionality.

\bibliography{bibliography}

\section*{Author contributions statement}

J.W. and A.P. conceived, developed, and implemented methods, interpreted results, and project managed. A.C. and Y.T. conceived and developed experiments, conducted experiments, and interpreted results. All authors contributed to the drafting and review of the manuscript.

\end{document}